\documentclass[aps,pra,twocolumn,showpacs,superscriptaddress]{revtex4}
\usepackage{amssymb,amsthm,amsmath,amsfonts}
\usepackage{color,hyperref,graphicx,ulem,subfigure}

\begin{document}
	\title{Balanced homodyne detection with on-off detector systems:\\Observable nonclassicality criteria}

	\author{J. Sperling}\affiliation{Arbeitsgruppe Theoretische Quantenoptik, Institut f\"ur Physik, Universit\"at Rostock, D-18051 Rostock, Germany}
	\author{W. Vogel}\affiliation{Arbeitsgruppe Theoretische Quantenoptik, Institut f\"ur Physik, Universit\"at Rostock, D-18051 Rostock, Germany}
	\author{G. S. Agarwal}\affiliation{Department of Physics, Oklahoma State University, Stillwater, Oklahoma 74078, USA}

	\date{\today}

\begin{abstract}
	Driven by single photon detection requirements, the theory of arrays of off-on detectors has been developed and applied.
	However for a comprehensive characterization of nonclassicality one also needs phase sensitive properties.
	This missing link is introduced with the derived theory of phase sensitive click counting measurements.
	It unifies the balanced homodyne detection for high intensities with the click detection in the few photon regime.
	We formulate a hierarchy of nonlinear squeezing conditions to probe quantum effects beyond standard squeezing.
	Imperfections stemming from fluctuations, detector efficiency, and dark count rates are considered.
	Experimentally accessible sampling formulas are given.
	Our theory paves the way towards novel applications of light in quantum metrology. 
\end{abstract}

\maketitle

\section{Introduction}

	Interference plays a crucial role for both quantum physics and classical wave theories.
	In quantum optics quantum interferences and superimposed electromagnetic field components occur simultaneously.
	Therefore, a proper analysis and determination of the character of interference patterns is indispensable for separating classical wave phenomena from quantum effects.

\begin{figure*}[ht!]
		\centering
		\subfigure[\label{fig:setup}]{\includegraphics*[width=5cm]{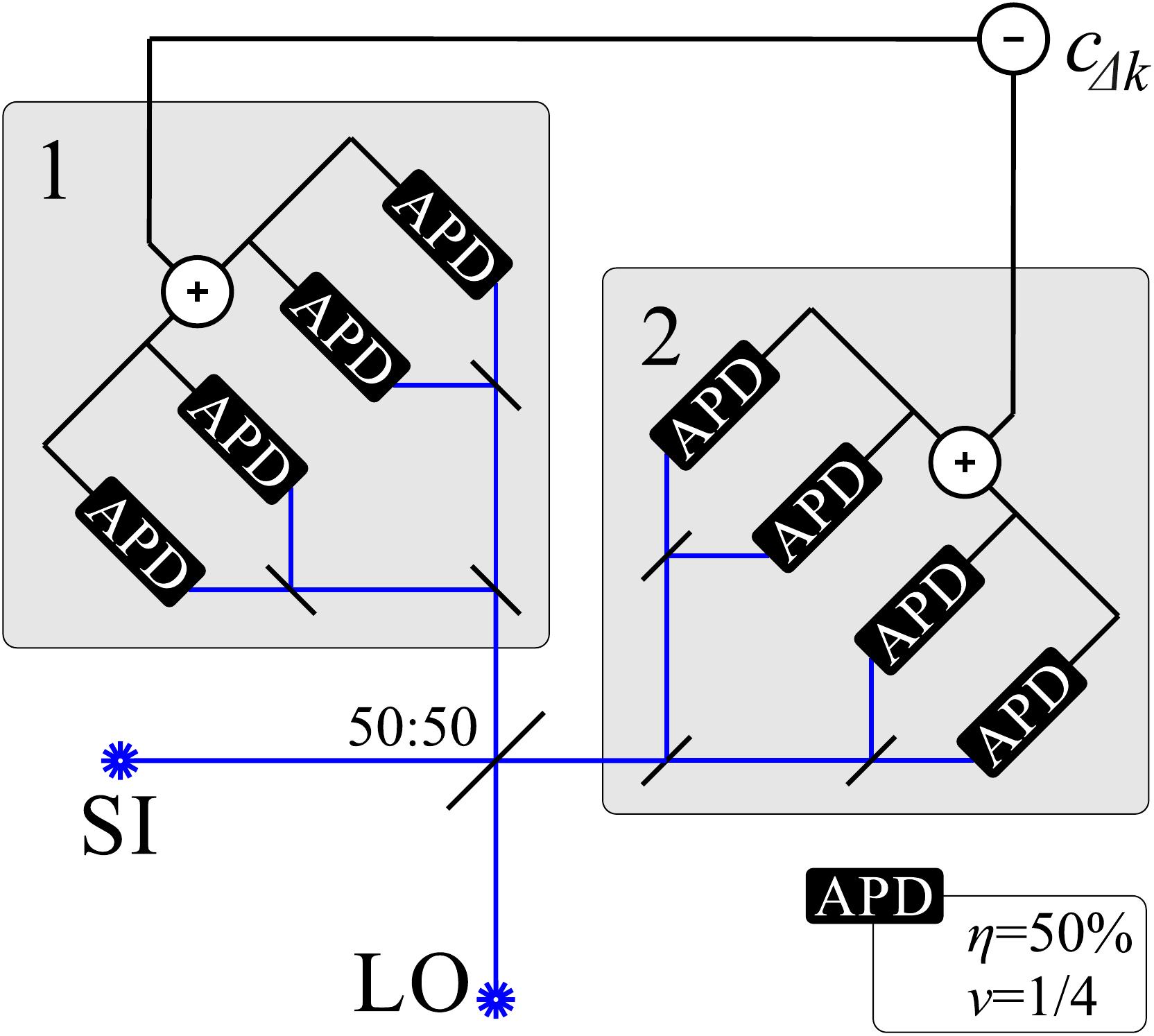}}\hspace*{0.25cm}
		\subfigure[\label{fig:clickstatistics}]{\includegraphics*[width=5.5cm]{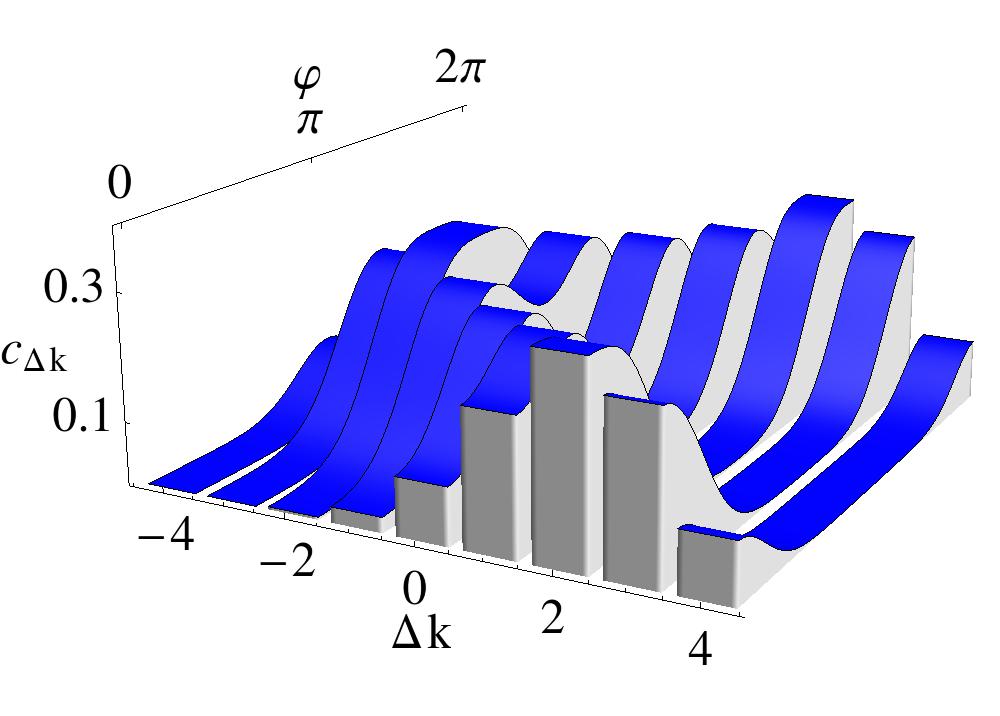}}\hspace*{0.25cm}
		\subfigure[\label{fig:nonlinquadrature}]{\includegraphics*[width=5.5cm]{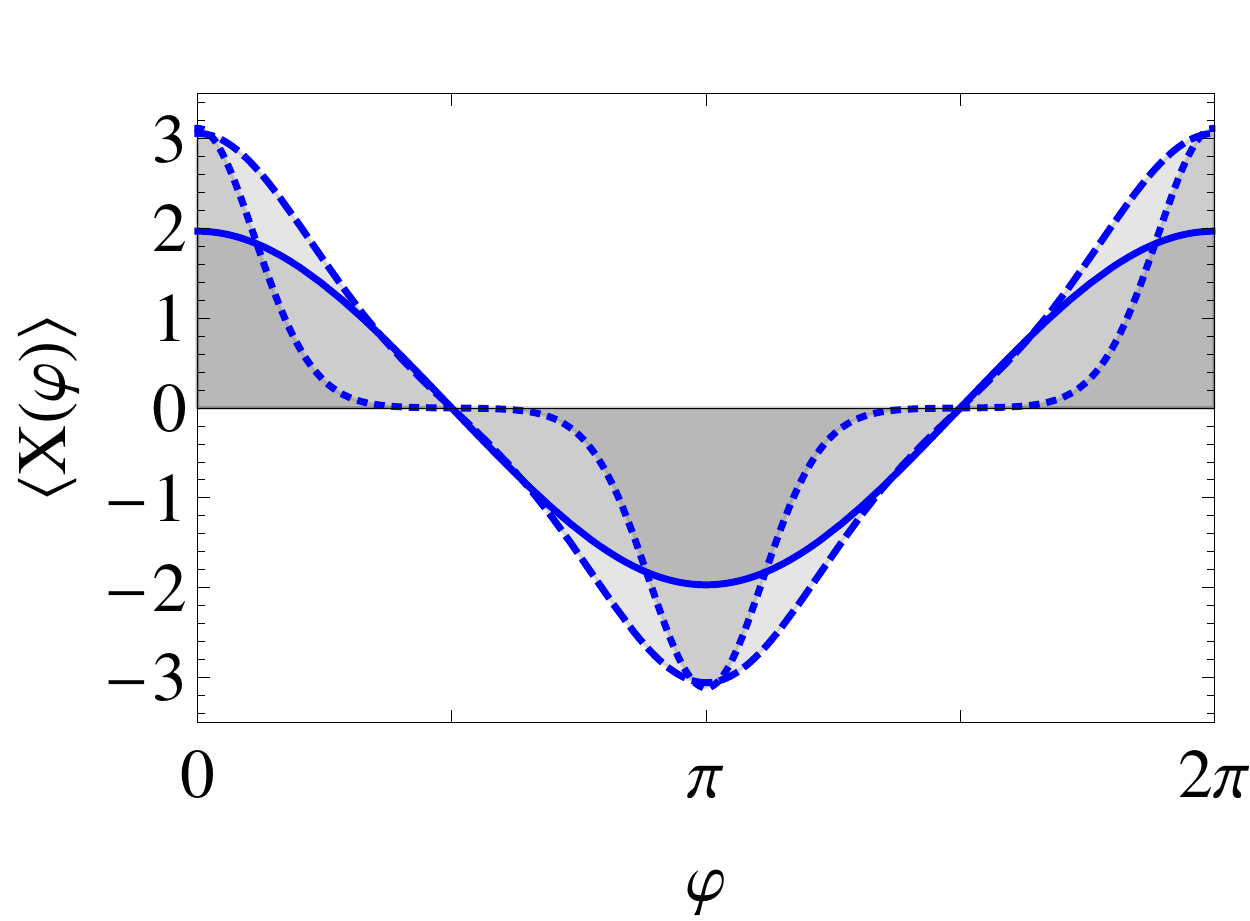}}
		\caption{(Color online)
			Figure~\ref{fig:setup} shows the setup of BHD-type of measurements with click counters employing $N=4$ APDs; here in a spatial-multiplexing configuration.
			The difference of clicks yields the difference count rate $c_{\Delta k}$, eq.~\eqref{eq:differencecountstatistics}, which is shown in~\ref{fig:clickstatistics} for a coherent state $|\alpha\rangle$ ($\alpha=2$) depending on the phase of the LO, $\beta=r{\rm e}^{{\rm i}\varphi}$ ($r=2$).
			Figure~\ref{fig:nonlinquadrature} depicts eq.~\eqref{eq:nonlinearquadratureoperator-expectation} for a coherent SI state and for the LO at different intensities, $r=\alpha=2,4,8$ (solid, dashed, dotted).
		}\label{fig:BHDsetup}
	\end{figure*}

	As photons reflect the particle nature of the electromagnetic field, their generation and detection are of fundamental interest~\cite{BJML10,SCSWS11,HSRHMSS14}.
	In the low photon number regime, detectors are often based on avalanche photodiodes (APD) in the Geiger mode.
	APDs produce a ``click'' for any number of absorbed photons and remain silent otherwise, {\it i.e.} ``no-click''.
	The incident light field can be split into fields with equal intensities, each being measured with an APD, to extract information beyond the binary one.
	Realizations are time- and spatial-multiplexed detectors and equally illuminated arrays of APDs; {\it cf.}~\cite{ALCS10,ZCDPLJSPW12} for recent experiments.
	In fig.~\ref{fig:setup}, two spatial-multiplexing schemes -- labeled with ``1'' and ``2'' -- are considered.
	Click counting devices have been applied in quantum metrology~\cite{XHP13,PSKHB07,BDFL08,DZTDSW11}, or for the determination of entanglement~\cite{MDL14,KWM00,PDFEPW10}.
	Recently, weak-field homodyning with multiplexing detectors has been implemented to probe quantum features of light~\cite{LCGS10,DBJVDBW14}.

	The click counting statistics $c_{k}$ for a measurement with $N$ APDs is of the form
	\begin{align}\label{eq:clickcountingstatistics}
		c_{k}=\left\langle {:}
			\binom{N}{k}\hat \pi^{k}(\hat 1-\hat \pi)^{N-k}
		{:}\right\rangle,
	\end{align}
	where $k$ is the number of clicks, $N$ the number of APDs, and ${:}\,\cdot\,{:}$ the normal ordering prescription~\cite{ClickTheo}.
	This quantum version of a binomial statistics is described by the operator $\hat\pi=\hat 1-{:}\exp[-(\eta\hat n/N+\nu)]{:}$, whose expectation value is the probability to record a click with a single APD.
        Herein $\hat n$ is the photon number operator, $\eta$ the quantum efficiency, and $\nu$ the dark count rate of each APD.
	Quantum properties, {\it e.g.} sub-binomial light, can be verified with such detection schemes~\cite{ClickQB,BDJDBW13,C14}.
	Moreover, higher order correlations, nonlinear absorption processes, multi-time correlations, and state engineering protocols have been investigated for these click counters~\cite{ClickCorr,ClickEngin}.

	In the high intensity regime, {\it i.e.} a few-photon approximation is not valid, the wave nature of the electromagnetic field is often studied by balanced homodyne detection (BHD) which is based on the photoelectric detection theory~\cite{WVO99,LR09,A13}.
	A signal (SI) is mixed on a beam splitter with a much stronger local oscillator (LO) while controlling the relative phase.
	Interferometric measurements, such as BHD, have been used to verify phase dependent nonclassical phenomena~\cite{KVHDSS09}.
	In this scenario, one would typically not apply single-photon counters because the total intensity of LO and SI is considered to exceed the capabilities of click detection.
	Instead, the outputs fields are measured with a detector which produces an electric current being proportional to the intensity of the incident light field.
	Applications have been established in quantum metrology~\cite{PADLA10,CIDE14} and weak LO homodyning~\cite{VG93,KWM00,PDFEPW10}.

	A nonclassical light field may be characterized by the non-existence a positive semi-definite Glauber-Sudarshan $P$~representation~\cite{GS63}.
	Moment based criteria in terms of field quadratures have been derived to verify such quantum correlations using BHD~\cite{HM85,H87,A93,SRV05}.
	Nonclassical light fields characterized in such a form, such as squeezed states, serve as a fundamental resource in applications which require a superior phase determination, for example, for gravitational wave detectors~\cite{C81,LIGO11}.
	In order to verify other correlations, {\it e.g.}, entanglement, it is indispensable to perform a careful detector analysis~\cite{SV11}.
	A phase sensitive detection theory does not exist for click counting devices, so that phase sensitive effects are nonaccessible in this regime.

	In the present contribution we formulate a BHD-type phase resolving click counting theory, which unifies the detection of the particle and the wave nature of quantum radiation fields.
	This yields a new observable -- a nonlinear quadrature operator -- as the key element of our approach.
	We provide a hierarchy of nonclassicality criteria to uncover phase sensitive quantum effects using on-off detector systems.
	Direct sampling formulas are given which allow to implement our theory in experiments.

\section{Measurement setup and nonlinear quadrature}

	In fig.~\ref{fig:setup} we describe the BHD setup using click counting devices.
	The SI field and the LO are combined on a 50:50 beam splitter.
	We assume that the LO and the SI are single mode fields with a perfect overlap.
	For the time being, we also say that the LO is in a perfect coherent state given by the coherent amplitude $\beta=r{\rm e}^{{\rm i}\varphi}$.
	The output beams are individually detected with a system of on-off detectors, which yields a joint click counting statistics $c_{k_1,k_2}$~\cite{ClickCorr}.
	For simplicity, we assume that both detector systems, 1 and 2, have identical characteristics, {\it i.e.}, number of APDs $N_1=N_2=N$, quantum efficiencies $\eta_1=\eta_2=\eta$, and dark count rates $\nu_1=\nu_2=\nu$.
	Finally, the difference count rate $c_{\Delta k}$ can be obtained as
	\begin{align}\label{eq:differencecountstatistics}
		c_{\Delta k}=\sum_{k_1-k_2=\Delta k} c_{k_1,k_2}.
	\end{align}
	This difference click counting statistics -- depending on the phase of the LO, $c_{\Delta k}=c_{\Delta k}(\varphi)$ -- is shown for a coherent SI state $|\alpha\rangle$ in fig.~\ref{fig:clickstatistics}.
	Note that a low intensity approximation with photoelectric detectors, having a Poissonian form of statistics, is impossible in this case since the mean number of photons from SI and LO is in the same order as the number of diodes~\cite{ClickTheo}.
	In general, we will refer to a high intensity if the total intensity of LO and SI exceeds the total number of APDs, $2N$.
	In this high intensity regime, the click counting statistics~\eqref{eq:differencecountstatistics} is valid even if the probability to have multiple photons at one detector significantly differs from zero.

	Now, the question arises how to infer quantum properties from this measurement.
	Since moment-based criteria turn out to be a fruitful approach, we may initially define a nonlinear quadrature operator $\hat X(\varphi)$ through
	\begin{align}\label{eq:nonlinearquadratureoperator}
		\langle\hat X(\varphi)\rangle=\langle N(\hat\pi_1-\hat \pi_2)\rangle= \sum_{\Delta k=-N}^N c_{\Delta k} \Delta k,
	\end{align}
	where $\hat \pi_{1(2)}=\hat1-{:}\exp[-(\eta\hat n_{1(2)}/N+\nu)]{:}$ describe the first(second) detector, respectively.
	Normally ordered powers of the operator $\hat X(\varphi)$ will be used later to uncover quantum effects.
	Using the beam splitter transform, $\hat a_1=(\hat a_{\rm SI}+\hat a_{\rm LO})/\sqrt 2$ and $\hat a_2=(\hat a_{\rm SI}-\hat a_{\rm LO})/\sqrt 2$, we rewrite eq.~\eqref{eq:nonlinearquadratureoperator} for the coherent LO, $\beta=r{\rm e}^{{\rm i}\varphi}$, as
	\begin{align}\label{eq:nonlinearquadratureoperator-expectation}
		\nonumber \langle\hat X(\varphi)\rangle{=}&N\left\langle\!{:}\!
			\exp\!\left[-\!\left(\!\frac{\eta}{N}\hat a_2^\dagger\hat a_2{+}\nu\!\right)\!\right]\!\!
			{-}\exp\!\left[-\!\left(\!\frac{\eta}{N}\hat a_1^\dagger\hat a_1{+}\nu\!\right)\!\right]
		\!{:}\!\right\rangle
		\\{=}&2 N{\rm e}^{-\frac{\eta r^2}{2N}-\nu} \left\langle\!{:}\,{\rm e}^{-\frac{\eta}{2N}\hat n}\sinh\left[\frac{\eta r}{2N}\hat x(\varphi)\right]{:}\!\right\rangle,
	\end{align}
	with the photon number $\hat n=\hat a^\dagger_{\rm SI}\hat a_{\rm SI}$ and the linear quadrature $\hat x(\varphi)=\hat a_{\rm SI}{\rm e}^{-{\rm i}\varphi}+\hat a^\dagger_{\rm SI}{\rm e}^{{\rm i}\varphi}$ of the signal field.

	We observe several features of the click quadrature operator $\hat X(\varphi)$.
	First, it has the intensity dependent contribution, $\exp[-\eta \hat n/(2N)]$, which is limiting the range of possible expectation values.
	This makes sense because for any SI and LO power the expectation value must not exceed the values $\pm N$, {\it cf.} the right hand side in~\eqref{eq:nonlinearquadratureoperator}.
	Second, the quadrature $\hat X(\varphi)$ is nonlinearly related to the SI's true quadrature $\hat x(\varphi)$ in terms of a hyperbolic sine.
	In fig.~\ref{fig:nonlinquadrature}, we plot the expectation value~\eqref{eq:nonlinearquadratureoperator-expectation} for a coherent SI with different intensities to show this nonlinear behavior.
	Finally we can analyze the limit $N\to\infty$,
	\begin{align}
		\lim_{N\to\infty}\langle\hat X(\varphi)\rangle={\rm e}^{-\nu}\eta r\langle{:}\hat x(\varphi){:}\rangle\propto\langle\hat x(\varphi)\rangle,
	\end{align}
	which corresponds to the linear quadrature being the result of BHD with photoelectric detection theory.
	Alternatively, a first order Taylor expansion in terms of low efficiencies $\eta\ll 1$ yields the same result.
	Moreover, the second order term includes intensity field correlations, $\langle{:}\hat n\hat x(\varphi){:}\rangle$, which have been studied in the context of nonclassicality determination~\cite{CO,V08}.
	The fact that we can get these number-amplitude cross correlation highlights the nonlinear character of $\hat X(\varphi)$ in eq.~\eqref{eq:nonlinearquadratureoperator-expectation}.
      	
\begin{figure}[ht!]
		\centering
		\includegraphics*[width=6.5cm]{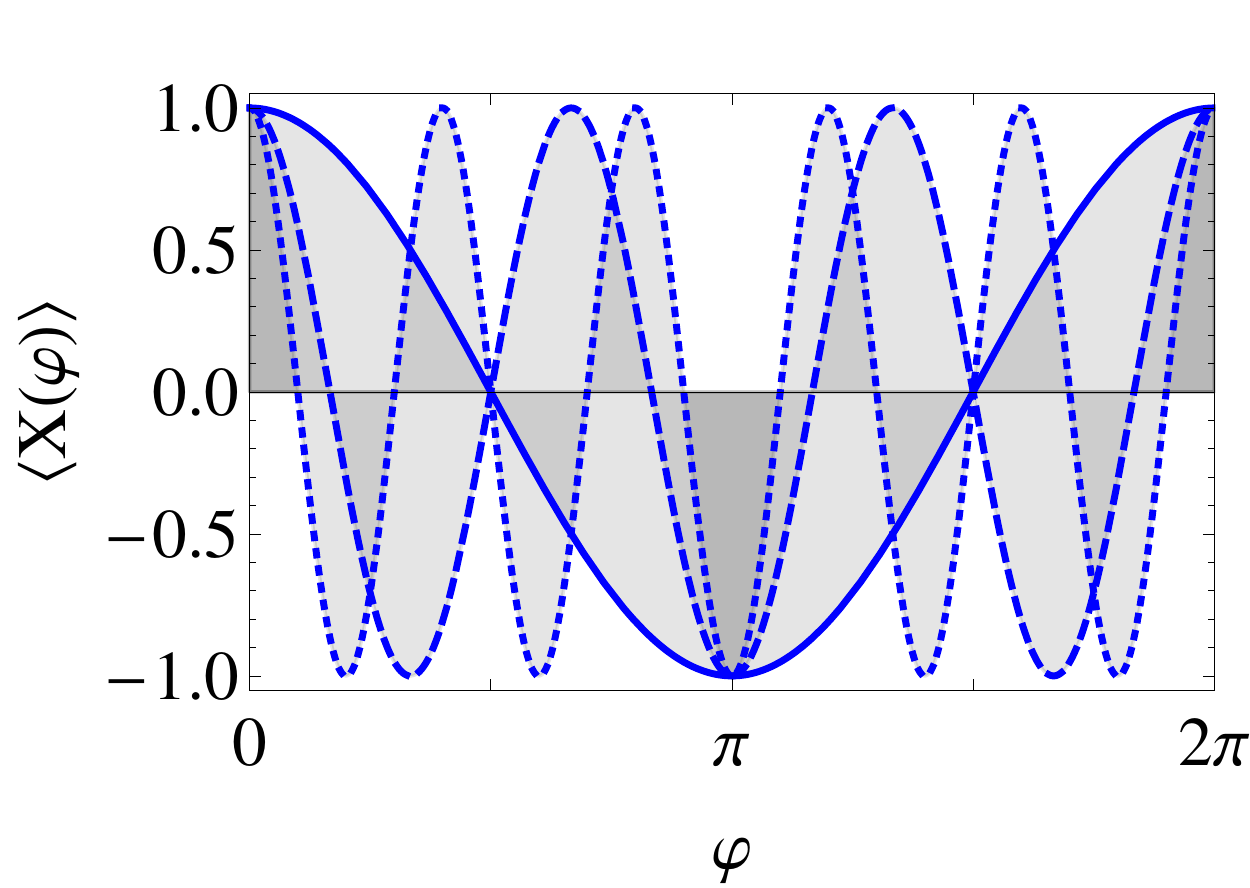}
		\caption{(Color online)
			The expectation value $\langle \hat X(\varphi)\rangle$ is shown for the states in eq.~\eqref{eq:singlemodeN00N}, $n=1,3,5$ (solid, dashed, dotted).
			For a better comparison the curves are scaled with $1/|\langle \hat X(0)\rangle|$: $10^{0.2},10^{2.4},10^{4.9}$ for $n=1,3,5$, respectively.
			The periods for increasing $n$ reduce to $2\pi/n$ for odd $n$.
			The detection parameters are $N=4$, $\eta=50\%$, $\nu=1/4$, and $r=2$.
		}\label{fig:photonquadrature}
	\end{figure}

	Let us consider a nonclassical signal of $n$-photons superimposed with vacuum,
	\begin{align}\label{eq:singlemodeN00N}
		|0{:}n\rangle=(|0\rangle+|n\rangle)/\sqrt 2,
	\end{align}
	being the single mode version of the so-called N00N state~\cite{BKABWD00}.
	Figure~\ref{fig:photonquadrature} shows the phase dependency of these states, $n=1,3,5$, with mean photon numbers $\langle \hat n\rangle=n/2$.
	The periodicity depends on the number of photons, which is of a particular interest for quantum metrology~\cite{BKABWD00,PADLA10}.

\section{Higher-order moments and nonclassicality}

	In order to formulate nonclassicality conditions, we may define the matrix of click quadrature operator moments,
	\begin{align}\label{eq:matrixofmoments}
		M=(\langle{:}\hat X^{m+m'}(\varphi){:}\rangle)_{m,m'=0}^{\lfloor N/2\rfloor},
	\end{align}
	with $\lfloor\,\cdot\,\rfloor$ being the floor function.
	Since this matrix of moments is formulated in terms of normally ordered operator powers, it has to be positive semi-definite for any classical light field; {\it cf.}, {\it e.g.},~\cite{ClickCorr}.
	Therefore, we can formulate sufficient nonclassicality criteria as follows.
	A light field is nonclassical, if for a choice of indices holds that
	\begin{align}\label{eq:nonclassicalityconditions}
		\det\left[(\langle{:}\hat X^{m+m'}(\varphi){:}\rangle)_{m,m'\in\mathcal I}\right]<0,
	\end{align}
	where the set $\mathcal I\subset\{0,\dots,\lfloor N/2\rfloor\}$ describes the rows and columns of the considered minor.
	For states with a positive semi-definite Glauber-Sudarshan $P$~functions, these minors are necessarily non-negative.
	Note that $\langle{:}\hat X^{0}(\varphi){:}\rangle=\langle \hat 1\rangle=1$ and $\langle{:}\hat X^{1}(\varphi){:}\rangle=\langle\hat X(\varphi)\rangle$.

	To guarantee the direct applicability in experiments, we formulate sampling formulas to determine the moments $\langle{:}\hat X^{m}(\varphi){:}\rangle$ directly from the measured joint click counting statistics $c_{k_1,k_2}=c_{k_1,k_2}(\varphi)$.
	Using the generating function approach in ref.~\cite{ClickCorr},
	\begin{align*}
		\langle{:}\hat\pi_1^{j_1}\hat\pi_2^{j_2}{:}\rangle=&\sum_{k_1=j_1}^N\sum_{k_2=j_2}^N\frac{k_1!k_2!(N-j_1)!(N-j_2)!}{N!^2(k_1-j_1)!(k_2-j_2)!}c_{k_1,k_2},
	\end{align*}
	we obtain
	\begin{align}
		\nonumber &\langle{:}\hat X^{m}(\varphi){:}\rangle=\langle{:}N^m[\hat \pi_1-\hat\pi_2]^m{:}\rangle
		\\ =&\sum_{j=0}^m\binom{m}{j}(-1)^{m-j}N^m\langle {:}\hat\pi_1^j\hat\pi_2^{m-j}{:}\rangle
		\\\nonumber=&\sum_{j=0}^m\sum_{k_1=j}^N\sum_{k_2=m-j}^N(-1)^{m-j}N^m\frac{\binom{m}{j}\binom{k_1}{j}\binom{k_2}{m-j}}{\binom{N}{j}\binom{N}{m-j}} c_{k_1,k_2}.
	\end{align}
	This renders it possible to certify nonclassical effects from the experimentally obtained counting statistics without time consuming data post processing.

	Because of its relevance in physics, we select from the hierarchy of nonclassicality conditions~\eqref{eq:nonclassicalityconditions} the second order one.
	For this case, $\mathcal I=\{0,1\}$, we get
	\begin{align}\label{eq:varianceconditions}
		0>\det\begin{pmatrix}
			1 & \langle{:}\hat X(\varphi){:}\rangle \\ \langle{:}\hat X(\varphi){:}\rangle & \langle{:}[\hat X(\varphi)]^2{:}\rangle
		\end{pmatrix}=\langle{:}[\Delta\hat X(\varphi)]^2{:}\rangle.
	\end{align}
	The vacuum field yields $\langle{:}\hat X(\varphi){:}\rangle_{\rm vac}=\langle{:}\hat X^2(\varphi){:}\rangle_{\rm vac}=0$.
	Hence, inequality~\eqref{eq:varianceconditions} can be formally expressed in the same form as in the case of standard BHD, {\it i.e.}: {\it Whenever the variance is below the variance of vacuum fluctuations, we have a nonclassical light field}; or, equivalently, $\langle{:}[\Delta\hat X(\varphi)]^2{:}\rangle<\langle{:}[\Delta\hat X(\varphi)]^2{:}\rangle_{\rm vac}$.

	\begin{figure}[ht!]
		\centering
		\includegraphics*[width=8cm]{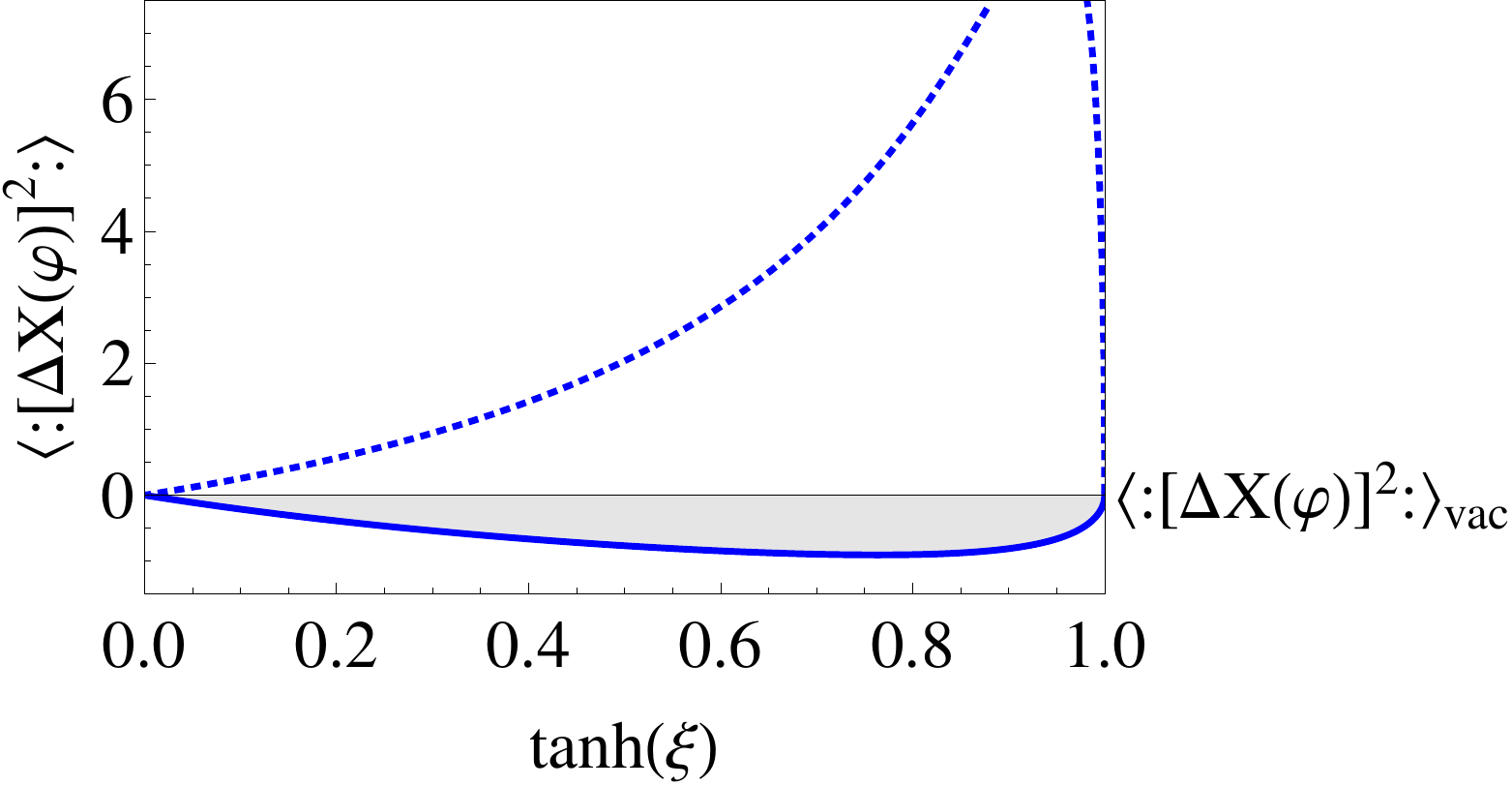}
		\caption{(Color online)
			The squeezed (solid) and anti-squeezed (dashed) nonlinear quadrature variance is given as a function of the squeezing parameter, $0<\xi<\infty$.
			Vacuum fluctuations yield the bound $\langle{:}[\Delta\hat X(\varphi)]^2{:}\rangle_{\rm vac}=0$.
			The detector characteristics is $N=4$, $\eta=50\%$, and $\nu=1/4$.
			The LO intensity is fixed, $r=2$, whereas the SI intensity goes up as $\xi$ increases.
			Eventually, the signal intensity even exceeds the LO intensity.
			Still, nonlinear squeezing can be observed.
		}\label{fig:variancesqueezed}
	\end{figure}

	As an example, we study a squeezed vacuum SI state,
	\begin{align}\label{eq:sqeezedvacuumstate}
		|\xi\rangle=\frac{1}{\sqrt{\cosh\xi}}\exp\left[-\frac{\tanh\xi}{2} \hat a_{\rm SI}^\dagger{}^2 \right]|{\rm vac}\rangle
	\end{align}
	characterized by the squeezing parameter $\xi>0$.
	Figure~\ref{fig:variancesqueezed} shows the nonlinear quadrature variance for $\varphi=0$($\pi/2$) corresponding to the squeezed(anti-squeezed) quadrature, respectively.
	We observe that even in the case of imperfect detection and a small number of APDs, we can identify squeezing through condition~\eqref{eq:varianceconditions} for all parameters $\xi$.
	It is also worth mentioning that the minimal normally ordered variance is $-1$.
	Due to continuity, the verification of quantum features for $\eta=50\%$ implies that this nonclassicality probe is also applicable for efficiencies below this value.
	For instance, we identified a relatively small squeezing for all $\xi$ values even for $\eta=1\%$, with a minimum of $\langle{:}[\Delta\hat X(\varphi)]^2{:}\rangle_{\rm min}\approx -7\cdot10^{-4}$.
	Another interesting feature is a saturation effect in fig.~\ref{fig:variancesqueezed}, $\xi\to\infty$ yields $\langle{:}[\Delta\hat X(\varphi)]^2{:}\rangle=0$.
	This is due to the fact that the intensity in such a case is so high that all APDs click all the time, which can be also achieved with a strong coherent SI, $|\alpha\rangle$ with $|\alpha|\to\infty$, being a classical state.
	Hence the limits of strong coherent light and infinitively squeezed vacuum cannot be discriminated from each other.

	Figure~\ref{fig:variancesqueezed} suggests that bounds to the linear squeezing,
	\begin{align}
		\langle [\Delta\hat x(\varphi)]^2\rangle={\rm e}^{-2\xi}<1=\langle [\Delta\hat x(\varphi)]^2\rangle_{\rm vac},
	\end{align}
	might be obtained from the nonlinear squeezing value.
	At least in the considered case of the pure states~\eqref{eq:sqeezedvacuumstate}, we can use the plot to retrieve the parameter $\xi$ from the nonlinear squeezing and anti-squeezing.
	This enables us to indirectly infer the linear quadrature squeezing levels using click counters.
	However, the case of mixed states requires further studies.

\section{Influence of losses and LO fluctuations}

\begin{figure}[ht!]
		\centering
		\includegraphics*[width=8cm]{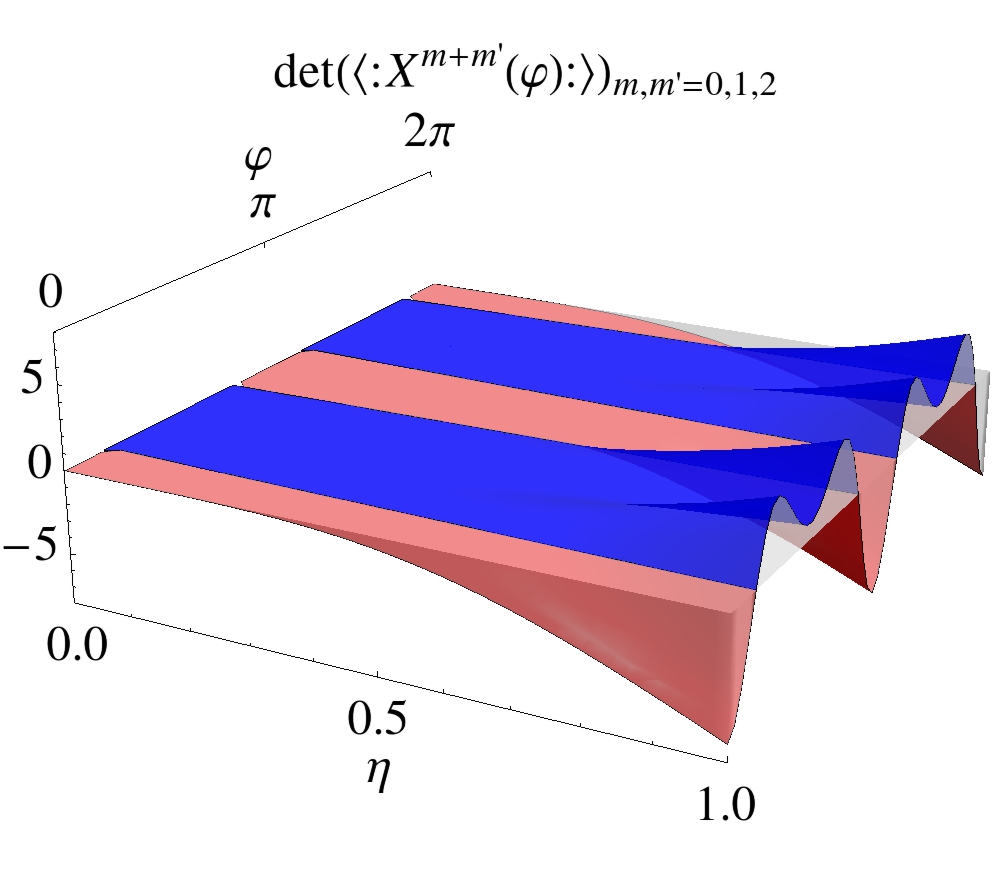}
		\caption{(Color online)
			The determinant of the matrix of moments~\eqref{eq:matrixofmoments} for the state state~\eqref{eq:singlemodeN00N}, $n=2$, is shown depending on the quantum efficiency $\eta$ and the phase $\varphi$.
			The other detection parameters are $\nu=1/4$, $r=2$, and $N=4$.
			Negativities (red) verify the nonclassicality, for some phases and all $\eta$~values.
		}\label{fig:highercorrelationefficiency}
	\end{figure}

	Another example is presented for the state in eq.~\eqref{eq:singlemodeN00N} using a higher-order nonclassicality condition~\eqref{eq:nonclassicalityconditions}; here $\mathcal I=\{0,1,2\}$.
	In fig.~\ref{fig:highercorrelationefficiency}, we plotted this minor depending on the phase $\varphi$ and the efficiency $\eta$.
	Since the variance of this state is non-negative, $\langle{:}[\Delta\hat X(\varphi)]^2{:}\rangle\geq0$, this scenario verifies that higher order correlations are useful to detect nonclassicality.
	Moreover, for all efficiencies, $1\geq \eta>0$, exist phase intervals with negativities which increase with increasing efficiency.

	So far we considered imperfections of the click detector itself, but not of the LO.
	In case of the standard BHD, classical LO fluctuation do basically not occur in the difference current of both detectors.
	Here such fluctuations can be crucial due to the nonlinear structure, {\it cf.} eq.~\eqref{eq:nonlinearquadratureoperator-expectation}.
	Thus we introduce a convolution of the unperturbed quadrature moments $\langle{:}\hat X^{m}(\varphi){:}\rangle$ with the LO noise distribution $P_{\rm LO}(\beta)$ and $\beta=r{\rm e}^{{\rm i}\varphi}$
	\begin{align}
		\langle{:}\hat X^{m}(\varphi){:}\rangle\mapsto\int {\rm d}^2\beta\, P_{\rm LO}(\beta)\langle{:}\hat X^{m}(\varphi){:}\rangle.
	\end{align}
	For a fundamental study, we infer Gaussian fluctuations, {\it cf.}~\cite{A87,FZ14}, being subdivided into phase noise $\sigma_\varphi$ and amplitude noise $\sigma_r$.
	The decomposition $\beta=(x+{\rm i}p){\rm e}^{{\rm i}\varphi}$ in a $\varphi$-rotated frame yields a Gaussian fluctuation distribution as
	\begin{align}
		P_{\rm LO}(\beta)=\frac{\exp\left[-\frac{(x-r)^2}{2\sigma_x^2}-\frac{p^2}{2\sigma_p^2}\right]}{2\pi\sigma_x\sigma_p},
	\end{align}
	having a mean coherent amplitude $\bar \beta=r{\rm e}^{{\rm i}\varphi}$.
	Using polar coordinates, the phase noise variance is given by $\sigma_p= r\sigma_{\varphi}$ and the amplitude noise variance by $\sigma_x=\sigma_r$.

\begin{figure}[ht!]
		\centering
		\includegraphics*[width=7.5cm]{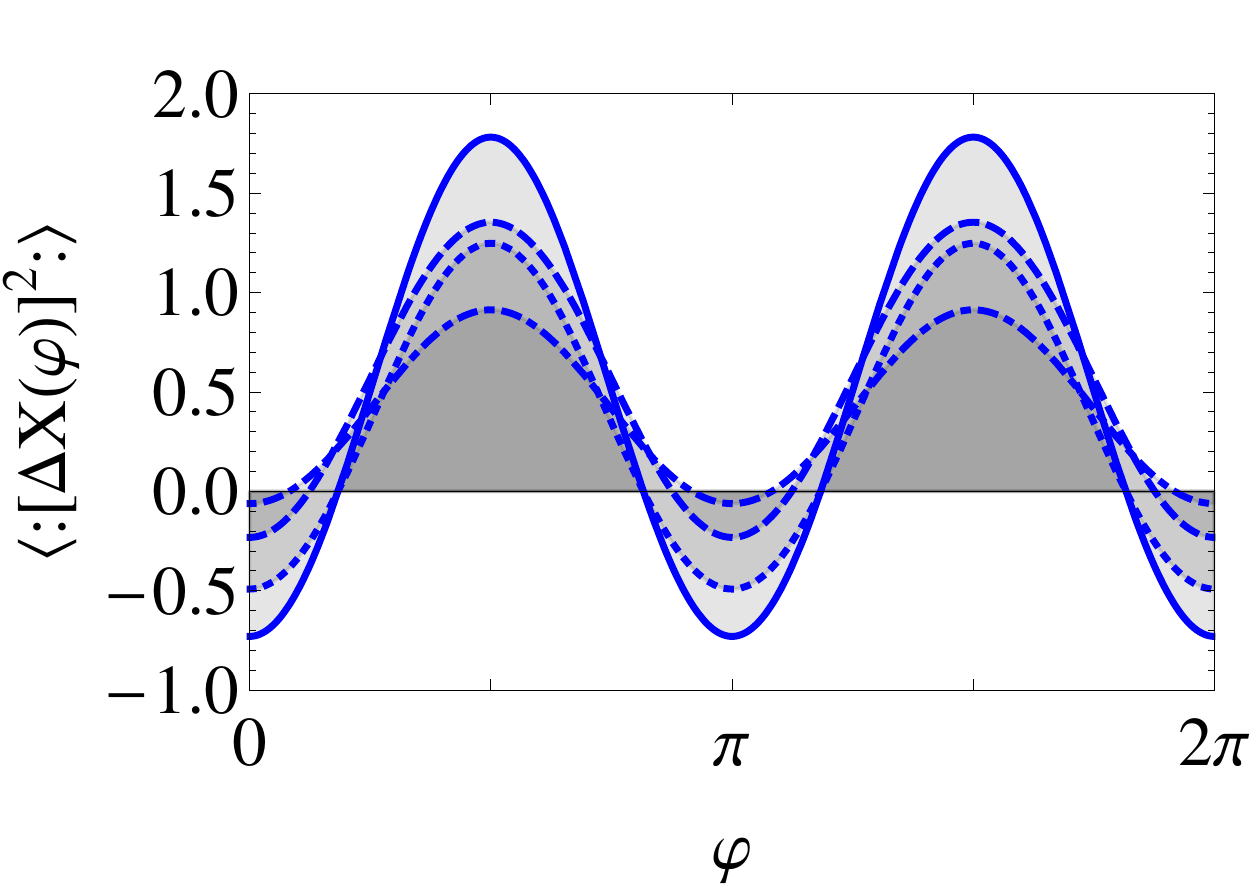}
		\caption{(Color online)
			The influence of LO noise to the nonlinear quadrature variance is shown -- with $N=4$, $\eta=50\%$, $\nu=1/4$, $r=2$, and $\xi=0.5$ -- for a squeezed vacuum state~\eqref{eq:sqeezedvacuumstate}.
			The solid line shows $\langle{:}[\Delta\hat X(\varphi)]^2{:}\rangle$ without LO fluctuations.
			The dashed graph includes a significant amount of phase noise, $\sigma_p=1.2$ ($\sigma_\varphi\approx34^\circ$), and the dotted one shows the effect of amplitude noise ($\sigma_x=\sigma_r=2$) of the same order of magnitude as the mean LO amplitude.
			The dot-dashed curve combines amplitude and phase noise.
		}\label{fig:variancephase}
	\end{figure}

	The influence of the LO fluctuations to the verification of nonlinear squeezing is shown in fig.~\ref{fig:variancephase}.
	The amplitude noise (dotted curve) solely affects the value of the negativities in comparison with the unperturbed variance (solid curve).
	The phase noise (dashed curve) additionally diminishes the intervals of certified squeezing.
	Typically, the phase is much better controlled than assumed for the plot.
	However, such high phase diffusion shall underline the applicability in extreme scenarios.
	Naturally both effects add up as shown by the dot-dashed curve.

	Let us outline the treatment of other possible sources of imperfections, which are not studied in detail in this work.
	Initially, we assumed a perfect mode overlap between the LO and the SI.
	However, a mode mismatch might occur in our scheme.
	In ref.~\cite{ClickCorr}, the characterization of the click counting devices has been generalized to a multimode description (section IV~B), which can be used to model an incomplete overlap between LO and SI.
	In the same work, time-dependent correlations in click detection have been elaborated (section IV~A).
	This technique is also useful to consider temporal drifts in the relative phase between LO and SI, or, in case of time-bin multiplexing, it renders it possible to treat temporal overlaps of wave-packets in different time bins.
	Based on these approaches, the influence of a manifold of experimental imperfection can be rigorously studied in addition to the perturbations presented above.

\section{Conclusions}

	Techniques for measuring the particle and the wave nature of quantum light have been unified in the theory of balanced homodyne detection with click counting devices.
	Its consistent formulation leads to a nonlinear quadrature operator as the basic observable.
	Its features have been studied for local oscillators in the weak and intermediate intensity regime, for detector imperfections, and fluctuations of the local oscillator.
	For consistency, we showed that standard balanced homodyne detection is recovered in proper limits.

	A hierarchy of conditions has been derived for determining phase sensitive nonclassical effects.
	The second order criterion applies to verify nonlinear squeezing of a squeezed vacuum state for arbitrary squeezing strength.
	Higher order criteria identify phase-sensitive nonclassicality of a superposition of $n$ photons with vacuum.
	Even in the case of quantum efficiencies below $50\%$, it has been demonstrated that nonclassicality can be uncovered in terms of higher-order correlations.
	By applying more than one phase sensitive click counting device, it is possible to infer phase sensitive correlations between multimode radiation fields.
	Our techniques can also be further developed for the aim of applications in quantum information and metrology.

\section*{Acknowledgements}
	JS and WV acknowledge financial support by Deutsche Forschungsgemeinschaft through SFB 652.

\end{document}